\documentclass[prd,aps,footinbib,longbibliography,twocolumn]{revtex4-2}
\usepackage[babel]{csquotes}
\usepackage{amsmath,amssymb}
\usepackage[colorlinks,citecolor=blue,linkcolor=blue,urlcolor=blue]{hyperref}
\usepackage{graphicx}
\usepackage{dcolumn}
\usepackage{bm}
\usepackage{physics} 
\usepackage{bbold}
\usepackage{mathtools}
\usepackage{xcolor}
\usepackage{tikz}
\usepackage{tikz-cd}
\usepackage{comment}
\usepackage{soul}
\usepackage{framed}
\usepackage{mdframed}
\usepackage{appendix}
\usepackage{hyperref}
\usepackage{natbib}
\usepackage[normalem]{ulem}
\usepackage[T1]{fontenc}
\usepackage{float}
\usepackage{amsthm}
\usepackage{caption}
\usepackage{subcaption}

\def\be{\begin{equation}}
\def\ee{\end{equation}}

\usepackage{xcolor}
\usepackage{xspace}
\usepackage{ifthen}

\usepackage[dvipsnames]{xcolor}

\newcommand{\cmmnt}[1]{}

 \newcommand{\showcomments}{false}

\newcommand{\marios}[1]%
{\ifthaenelse{\equal{\showcomments}{true}}%
{{\color{blue}{\small \textbf{MC:} #1}}}{\xspace}}%

\newcommand{\bw}[1]%
{\ifthenelse{\equal{\showcomments}{true}}%
{{\color{red}{ #1}}}
{\xspace}}%
\newcommand{\bwcomm}[1]%
{\ifthenelse{\equal{\showcomments}{true}}%
{{\color{blue}{ #1}}}
{\xspace}}%

\newcommand{\kai}[1]%
{\ifthenelse{\equal{\showcomments}{true}}%
{{\color{teal}{\small \textbf{KE:} #1}}}{\xspace}}%

\newcommand{\keith}[1]%
{\ifthenelse{\equal{\showcomments}{true}}%
{{\color{orange}{\small \textbf{KS:} #1}}}{\xspace}}%

\newcommand{\new}[1]
{{\color{black}{#1}}}{\xspace}%

\newcommand{\revise}[1]%
{{\color{blue}{#1}}}{\xspace}%

\usepackage{setspace} 
\usepackage[font=small,skip=4pt]{caption}

\begin{document}

\title{Quantum Sensing of Gravitational Frame-Dragging with a Superfluid $^4$He Gyrometer}

\author{Kai-Isaak Ellers$^{1,2,3}$, Marios Christodoulou$^{4}$, K.C. Schwab$^5$, K. Birgitta Whaley$^{2,3}$}
\affiliation{$^{1}$Department of Physics, University of California, Berkeley, CA, 94720, U.S.A.}
\affiliation{$^2$Department of Chemistry, University of California, Berkeley, CA, 94720, U.S.A.}
\affiliation{$^3$Challenge Institute for Quantum Computation, University of California, Berkeley, CA  94720}
\affiliation{$^{4}$Institute for Quantum Optics and Quantum Information (IQOQI), Austrian Academy of Sciences, Boltzmanngasse 3, A-1090 Vienna, Austria}
\affiliation{$^5$Thomas J. Watson, Sr., Laboratory of Applied Physics, California Institute of Technology, Pasadena, CA 91125, USA}

\date{\small\today}

\begin{abstract}

We propose a laboratory-scale experiment to locally measure the general relativistic frame-dragging effect on Earth using the macroscopic quantum properties of a novel superfluid $^4$He single Josephson junction gyrometer.  We derive the frame-dragging and related geodetic and Thomas effects in the superfluid gyrometer and present a procedure for their experimental measurement. We compute the expected thermal noise floor and find that very high sensitivity can be expected at millikelvin temperatures, where near-future Josephson junctions using nanoporous 2D materials are expected to operate. Assuming utilization of the lowest mechanical loss materials, we find a noise spectral density of $5\times 10^{-17}$ rads/s/$\sqrt{\mathrm{Hz}}$ at 10 mK, which is sufficient to resolve the frame-dragging rate to 0.2\% within one second of measurement, giving a rotational sensitivity of  1 revolution in 4 Byrs.  This extreme sensitivity to rotation corresponds to a measurement of proper time differences as small as $10^{-35}$ s. 

\end{abstract}

\maketitle

More than one hundred years after the first test of Einstein's theory of general relativity (GR)--Eddington's 1919 measurement of the deflection of light by the sun~\cite{dyson1920ix}--a great deal of effort is going towards testing the theory's predictions and searching for any deviations that might provide hints of new physics. The problems of dark matter and dark energy, as well as open questions about the quantization of gravity, may point to a breakdown of GR in some regimes. In particular, more precise measurements of frame-dragging and geodetic precession could test GR against general metric theories of gravity~\cite{will2014confrontation}, as well as Ho\v{r}ava-Lifshitz gravity~\cite{hovrava2009quantum,capozziello2021constraining} and Lorentz-violating standard model extensions~\cite{moseley2019lorentz}.

Frame-dragging due to the Earth's rotation was first measured by the LAGEOS satellites in 2004 to a level of 20\%~\cite{ciufolini2004confirmation}, and improved in 2019 to the 2\% level with the inclusion of the LARES satellite~\cite{ciufolini2019improved}. These estimates are reliant on complex modeling of the Earth's gravitational potential. Culminating in 2011, the Gravity Probe B satellite measured the effect to 18\% using four London moment gyroscopes, with the dominant source of error coming from torques on the gyroscopes due to electrical asphericity discovered after launch~\cite{everitt2011gravity}. Both of these measurements utilize classical probes to detect this small effect.

Local Earth-based measurements would avoid the need to average over many orbital periods and allow for experimental improvements. Here we propose a laboratory experiment capable of measuring frame-dragging at the 0.2\% level in one second of measurement
by harnessing the Josephson effect in superfluid $^4$He, together with ultra-low losses at millikelvin temperatures. This couples general relativity directly to a macroscopic quantum phenomenon. This is made possible due to recent advances in 2D nanomaterials, which are expected to realize Josephson effects in $^4$He for temperatures far below the superfluid transition temperature, $T_\lambda\sim 2.17K$. Towards the end of this letter, we compare with the GINGER proposal for an Earth-based measurement using a ring-laser gyroscope (RLG) \cite{bosi2011measuring}.

\begin{figure}[t]
\centering
    \begin{subfigure}[c]{0.22\textwidth}
        \centering
        \includegraphics[width=\textwidth]{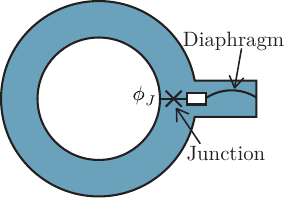}
        \label{fig:gyro}
    \end{subfigure}
    \hfill
    \begin{subfigure}[c]{0.21\textwidth}
        \includegraphics[width=0.9\textwidth]{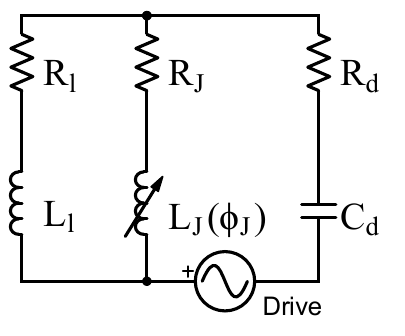}
        \label{fig:circuit}
    \end{subfigure}
    \caption{(a) Superfluid $^4$He gyrometer (left) consisting of a loop of superfluid $^4$He interrupted by a Josephson junction, connected in parallel with a controllable diaphragm to constitute a hydrodynamic Helmholtz resonator.  Rotation perturbs the quantum phase difference across the junction, thereby modifying the Helmholtz resonant frequency, $\omega_H$. (b) Equivalent electrical circuit (right) showing the hydrodynamic inductance of the loop $L_l$, the junction $L_J(\phi_J)$, the hydrodynamic capacitance $C_d$, and the resistive losses $R_l$, $R_J$, and $R_d$.}\label{fig:gyroscope}
\end{figure}

Superfluid helium demonstrates quantum coherence over macroscopic distances, frictionless motion, and quantized motion around loops. When placed in a SQUID-like arrangement (see Fig.~\ref{fig:gyroscope} (a)) and rotated, the resulting motion of the fluid will produce quantum phase-shifts across a barrier. These phases correspond to minute proper time dilation accumulated along the loop, due to the presence of off--diagonal components in the gravitational metric. A Josephson junction in this barrier can produce measurable changes in the superfluid transport, providing a way to detect the quantum phase.

Since the 1980s~\cite{anandan1981gravitational}, it has been proposed that superfluid SQUID-like devices could harness these properties to form ultra-sensitive gyrometers capable of detecting frame-dragging. In such systems, neutral helium atoms couple to gravity through their mass and sense the gravitomagnetic flux. Unlike conventional gyroscopes, rotation is measured via Aharonov–Bohm-type phases rather than the precession of a physical system, hence the term gyrometer.

Since the late 1990's a number of $^3$He~\cite{simmonds2001quantum,avenel2004superfluid} and $^4$He~\cite{schwab1997detection,hoskinson2006superfluid} gyrometers have demonstrated the expected coupling between quantum phase and rotation, but with noise levels many orders of magnitude higher than what is required to access frame-dragging. For $^3$He, the realization of a Josephson junction as a parallel array of weak-links has been straightforward due to the relatively large size of the coherence length, $\xi_0\sim 65$ nm~\cite{simmonds2001quantum}.  However, the sensitivity is ultimately limited by the difficulty of cooling the condensate far below the transition temperature $T_c\sim 1$ mK, so as to reduce the dissipation and accompanying thermal noise due to the  viscosity of the normal fluid fraction~\cite{avenel2004superfluid}.

For $^4$He, the challenge has been in the realization of a suitable Josephson junction that can function far below $T_\lambda\sim~2.17$ K. The coherence length in $^4$He for temperatures $T<T_\lambda$ scales as $\xi(T)=\xi_0 (1-T/T_\lambda)^{-2/3}$~\cite{duc2015critical}, with $\xi_0\sim 0.3$ nm at mK temperatures, which is essentially atomic size and outside the capability of conventional ``top-down'' nanofabrication. It has been possible to realize Josephson behavior if one approaches within 1 mK of $T_\lambda$, where the coherence length becomes the size of a nanofabricated aperture~\cite{sukhatme2001observation}.  However, here the relatively high temperature and large fraction of viscous normal fluid severely limit the ultimate thermal noise floor~\cite{chui2005frequency}.

Recent advances in atomically-engineered, synthetic 2D nanoporous materials appear well-suited to realize Josephson junctions for $^4$He made from extended arrays of $\sim$~1 nm apertures~\cite{kissel2014nanoporous}, which would allow operation at temperatures far below $T_\lambda$. Furthermore, it is possible to cool $^4$He very deep into the condensate phase, where $T/T_\lambda\approx  1/200$ at $T=10$ mK.  At these temperatures, dissipation due to the thermal excitations falls rapidly and coupling to containing structures has been shown to be very weak~\cite{de2017ultra}. These are the key reasons for the extreme sensitivity derived here for a $^4$He gyrometer, which would allow measurement of the frame-dragging effect to unprecedented precision.

\emph{Superfluid gyrometers ---} When cooled into the superfluid phase, $^4$He can be described by a macroscopic complex order parameter $\psi = \sqrt{\rho} e^{i\phi}$, where $\rho$ is the superfluid density and $\phi$ is a quantum mechanical phase \cite{anderson1966considerations}. When two regions of superfluid condensate are put into weak contact, the transport obeys the Josephson equations~\cite{anderson1966considerations}:
\begin{equation} \label{josephson}
            I = I_c \sin{\phi_J} ~~~~~~~~ \frac{d\phi_J}{dt} = -\frac{\Delta \mu}{\hbar}
\end{equation}
where $I$ is the superfluid mass current, $I_c$ is the critical mass current and $\phi_J$ and $\Delta \mu$ are the differences in phase and chemical potential, respectively, across the weak link. This behavior has been observed for both $^3$He and $^4$He \cite{avenel1988josephson, pereverzev1997quantum, sukhatme2001observation, davis2002superfluid}. Therefore, such weak links behave analogously to superconducting Josephson junctions.

Here, we analyze the single-junction superfluid $^4$He gyrometer shown in Fig.~\ref{fig:gyroscope} (a). This device behaves as a hydrodynamic Helmholtz resonator, analogous to the RLC electrical circuit shown in Fig.~\ref{fig:gyroscope} (b). The Josephson junction acts as a nonlinear inductor, with phase-dependent inductance $L_J(\phi_J) = \kappa_4 / (2\pi I_c \cos(\phi_J))$, where $\kappa_4 = h/m$ and $m$ is the mass of a $^4$He atom. The hydrodynamic inductance, $L_l$, and capacitance, $C_d$, are associated with kinetic energy in the fluid flow in the loop and potential energy storage by the diaphragm, respectively. The resistances derive from dissipation in the diaphragm motion ($R_d$), in the fluid flow through sensing loop ($R_l$), and junction ($R_J$). Rotational and gravitational effects manifest themselves as changes in the Josephson phase $\phi_J$, modifying the Josephson inductance, which results in a measurable change in the resonant frequency, $\omega_H$, of the device. This frequency shift is detected by driving and measuring the response of the diaphragm~\cite{avenel2004superfluid,chui2005frequency}.

\emph{Comoving reference frame ---} 
General relativistic gravitational effects in superfluid gyrometers are most easily calculated in the coordinates used in \cite{ashby1990geodetic}, in which the Lense-Thirring metric up to linear order in $\vec{R}$ takes the form,
\begin{equation}
    \label{metric}
    \begin{aligned}
        &g_{00} = -1+\frac{2 \vec{R} \cdot \vec{a}}{c^2} \, , \qquad g_{ij} = \delta_{ij} \, , \\ 
        &\vec{g} = -\frac{1}{c}(\vec{\Omega}_{G}+\vec{\Omega}_{T} + \vec{\Omega}_{FD}) \times \vec{R} + K \vec{R} \, ,
    \end{aligned}
\end{equation}
where $\vec{a}$ is the acceleration of the lab and $\vec{g} = (g_{01},g_{02},g_{03})$ is the gravitomagnetic vector potential. The three angular velocities $\vec{\Omega}$ are the precession rates due to the Lense-Thirring frame-dragging, the geodetic and the Thomas effects. They play the role of gravitomagnetic fields. The term $K\vec{R}$ does not contribute to the Josephson phase and can be removed by a coordinate transformation~\cite{ashby1990geodetic}. Expressions for $\vec{\Omega}$ and $K$ are given in Appendix~\ref{app:A}. The coordinates correspond to a frame comoving with the superfluid gyrometer but with the orientation of its spatial axes fixed with respect to distant stars. 

\emph{Gravitational effects in a superfluid gyrometer---}
Both rotational motion and the Earth's gravitational field will modify the phase across a Josephson junction.  For example, it was noted many years ago that the quantized flux through a superconducting ring includes contributions from both magnetic and gravitational (Lense-Thirring) fields~\cite{dewitt1966superconductors}. However, superconducting rings are not suitable as gyrometers because of the difficulties of canceling external magnetic fields~\cite{satterthwaite1997concerning}. 

For the case of $^4$He atoms in superfluid, which are uncharged, the point-particle Lagrangian is $L = -m (-g_{\mu\nu} \dot{x}^\mu \dot{x}^\nu )^{1/2}$, where $m$ is the atomic mass. The Hamiltonian is then $H = \frac{1}{2m} ( \vec{p} - mc \vec{g} )^2 - \frac{m}{2} h_{00}$, where $\vec{p}$ is the canonical momentum of the atom and $h_{00} = g_{00} + 1$ is the metric perturbation. Assuming constant density $\rho$, we find that the superfluid velocity is $\vec{v}_s = \frac{\hbar}{m} \nabla \phi - c \vec{g}$. For details see Appendix~\ref{app:A}.

To ensure that the wavefunction is well-defined, the phase must be single-valued at each point, up to integer multiples of $2\pi$. Thus, the phase change along any closed spatial path $\gamma_c$ through the superfluid is quantized: $\Delta \phi[\gamma_c] = 2\pi n$, for some integer $n$~\cite{tilley2019superfluidity}. Here we assume the zero-circulation state, which implies $n=0$. For a superfluid ring interrupted by a junction with phase $\phi_J$ across it, the phase change going around the loop once is then
\begin{equation} \label{circ_quant}
    \Delta \phi[\gamma_c] = \frac{m}{\hbar} \oint \vec{v}_s \cdot d\vec{R} + \frac{mc}{\hbar} \oint \vec{g} \cdot d\vec{R} + \phi_J = 0
\end{equation}

The junction phase will have both DC and oscillating components, $\phi_J(t)=\phi_0+\phi_A(t)$ (see Appendix~\ref{app:B}). The effects of rotation and gravity will contribute a small perturbation $\Delta \phi_0$ to the DC phase. The presence of the rigid partition containing the Josephson junction entrains the superfluid into motion that is well-approximated at some radius by the solid-body velocity (except near the junction)~\cite{packard1992principles}. Therefore, there exists a contour on which $\vec{v}_s \approx \vec{\Omega}_\oplus \times \vec{R}$, where $\vec{\Omega}_\oplus$ is the Earth's angular velocity. The two integrals in \eqref{circ_quant} can be evaluated using Stokes' theorem and the identity $\nabla \times (\vec{\Omega} \times \vec{R}) = 2\vec{\Omega} $, yielding
\begin{align} \label{phase_tot}
    \Delta \phi_0 & = -\frac{m}{\hbar} \oint \left(\vec{\Omega}_\oplus \times \vec{R} \right) \cdot d\vec{R} -\frac{m c}{\hbar} \oint \vec{g} \cdot d\vec{R}   \nonumber \\ 
    & = -\frac{2m}{\hbar} \vec{\Omega}_{tot} \cdot \vec{A} \, ,
\end{align}
where we defined $\vec{\Omega}_{tot} = \vec{\Omega}_\oplus - \vec{\Omega}_{FD} - \vec{\Omega}_G - \vec{\Omega}_T$ and where $\vec{A}$ is the area vector normal to the loop. 
We see that $\Delta \phi_0$ is measuring the total rotational flux through the loop. Note that the  $K\vec{R}$ term in the off--diagonal metric components $\vec{g}$, see \eqref{metric}, does not contribute to the loop integral, because it is curl-free. The term involving $\vec{\Omega}_\oplus$ in \eqref{phase_tot} yields the Sagnac phase~\cite{rizzi2003sagnac,ruggiero2015note,noteSagnac}, which does not depend on the gravitational constant $G$. The phase arising from the loop integral of $\vec{g}$ in \eqref{phase_tot} has the form of an Aharonov-Bohm phase, with $\vec{g}$ playing the role of a gravitomagnetic vector potential~\cite{chiao2014gravitational}. Its curl, $ \nabla \times \vec{g}= -2( \vec{\Omega}_{FD}+\vec{\Omega}_G + \vec{\Omega}_T)/c$, plays the role of a gravitomagnetic field~\cite{ciufolini1995gravitation}. 

\emph{Relation to proper time dilation ---} 
Quantum phases due to gravity can be interpreted as a proper time dilation $\delta \tau$, through the formula $\Delta \phi = mc^2 \delta \tau/\hbar$ ~\cite{christodoulou2019possibility,christodoulou2022experiment,rovelli2018possibility}. Here, the proper time dilation related to $\phi_J$ arises from the presence of cross-terms in the metric. These give rise to a difference in the proper time experienced by observers moving clockwise or counterclockwise around a closed spatial contour~\cite{hafele1972around,ruggiero2015note}. Then, clocks cannot be consistently synchronized along a closed spatial contour $\gamma_c$; there is a proper time difference $\delta \tau[\gamma_c] = -\frac{1}{c}\oint_{\gamma_c} \frac{g_{0i}}{\sqrt{-g_{00}}} dx^i$ upon returning to the starting point~\cite{landau2013classical}. Another way to see this is to note that $-\frac{1}{c} \oint \vec{g} \cdot d\vec{R}=-\frac{1}{c}\oint \vec{g} \cdot \vec{v} \, dt$ is precisely the proper time dilation due to the metric components $\vec{g} $ when linearizing $\tau = \int dt \sqrt{-g_{\mu\nu} \dot{x}^\mu \dot{x}^\nu}$ around the flat metric. Similarly, the Sagnac phase is the special relativistic contribution to time dilation due to rotation~\cite{rizzi2003sagnac}. Indeed, the phase derived in \eqref{phase_tot} is in correspondence with the phase difference found in \cite{bosi2011measuring} through a calculation of the proper time difference along the two counter-propagating beam paths in a ring laser arrangement. 

\begin{figure}
    \centering\includegraphics[width=0.2\textwidth]
    {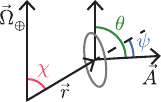}
    \caption{The angles $\theta$, $\chi$ and $\psi$ describe the relative orientations of the Earth angular velocity, $\vec{\Omega}_\oplus$, the gyrometer position relative to Earth center, $\vec{r}$, and the area vector normal to the gyrometer loop, $\vec{A}$.}
    \label{fig:angles}
\end{figure}

\emph{Experimental protocol ---} 
It is desirable to measure each rotational or gravitational effect independently, particularly for tests of GR against other metric theories of gravity. Let $\vec{r}$ be the vector from the Earth center to the position of the gyrometer in the parameterized post-Newtonian (PPN) frame (approximate rest-frame of the solar system, see Appendix \ref{app:A}). The combined rotational and gravitational contributions to the phase, $\Delta \phi_0$, can be expressed in terms of angles defined by $\hat{\Omega}_\oplus \cdot \hat{A} = \cos{\theta}$, $\hat{\Omega}_\oplus \cdot \hat{r} = \cos{\chi}$ and $\hat{r} \cdot \hat{A} = \cos{\psi}$, (see Fig. \ref{fig:angles}). For a gyrometer on Earth's surface, this yields $\Delta \phi_0 = \Delta \phi_{S} + \Delta \phi_{FD} + \Delta \phi_G + \Delta \phi_T$, with
\begin{equation} \label{phase_sagnac2}
    \Delta \phi_S = - \Phi_\oplus\ \cos{\theta}
\end{equation}
where $\Phi_\oplus = 2m\Omega_\oplus A/\hbar$. The $G$-dependent phases include the Newtonian potential $U_\oplus = GM_\oplus/R_\oplus$:
\begin{align} \label{eq:phases}
    \Delta \phi_{FD} &= \Phi_\oplus \, U_\oplus \frac{(1+\gamma+\frac{1}{4} \alpha_1)}{5 c^2 } \left(3 \cos{\chi} \cos{\psi} - \cos{\theta} \right) \nonumber \\
    \Delta \phi_G &= \Phi_\oplus \, U_\oplus \frac{(2\gamma+1)}{ 2c^2} \left(\cos{\theta} - \cos{\chi} \cos{\psi} \right), \nonumber \\
    \Delta \phi_T &= \Phi_\oplus \, U_\oplus \frac{1}{ 2 c^2} \left(\cos{\theta} - \cos{\chi} \cos{\psi} \right) \nonumber \\ &- \frac{m\Omega_\oplus^3 R_\oplus^2 A }{\hbar c^2} \sin^2{\chi} \cos{\theta}.
\end{align}

Here, $\gamma$ and $\alpha_1$ are post-Newtonian parameters that parameterize deviations from GR in general metric theories of gravity, with $\gamma=1$ and $\alpha_1=0$ in GR. All three phases $\Delta \phi_{FD}$, $\Delta \phi_G$ and $\Delta \phi_T$ are extremely small compared to the Sagnac phase $\Delta \phi_S$. The frame-dragging and geodetic phases, and the first term in the Thomas phase are $\sim$~5~$\times10^{-10} \Delta \phi_S$ while the second part of the Thomas phase is another two orders of magnitude smaller.
 The physical origin of these phases is as follows. Frame-dragging and geodetic effects arise from general relativity: the former results from Earth’s rotation, while the latter exists even for a static mass. The Thomas phase originates from non-gravitational forces, with the first term due to the normal force opposing gravity and the second from Earth’s rotation.

The Sagnac effect can be canceled by orienting the superfluid loop perpendicular to the axis of the Earth's rotation, $\vec{\Omega}_\oplus$, see Fig.~\ref{fig:angles}. Overcoming the $5\times 10^{10}$ difference in magnitude between the Sagnac and gravitational phases, requires control of the orientation of the gyrometer at the sub-nanoradian level. This is the same order of magnitude that was required in the control of the satellite orientation relative to a guide star for Gravity Probe B~\cite{everitt2015gravity}. Ring-laser gyrometer experiments have also demonstrated angular control at the nanoradian level using tiltmeters~\cite{schreiber2023variations}. The angular control will also have to accommodate variation in the orientation of the Earth's rotational axis relative to fixed stars (tracked by the International Earth Rotation and Reference Systems Service (IERS) with a precision of $0.1$ mas~\cite{bosi2011measuring} or $4.8 \times 10^{-10}$ rad).

\emph{Thermal noise limit on rotational sensitivity ---} The rotational sensitivity of a superfluid gyrometer is fundamentally limited by thermal noise originating from the dissipative elements in the circuit~\cite{avenel2004superfluid}. Previous work has analyzed the resolution of superfluid He gyrometers due to noise generated by the viscous normal fluid~\cite{chui2005frequency}.  
This is appropriate for superfluid $^3$He  when $T/T_c\gtrsim 0.2 $~\cite{greywall1984thermal,vollhardt2013superfluid}, and for $^4$He above $\sim$~0.7 K. However, for $^4$He below $\sim 0.6$ K, the mean free path of the thermal excitations becomes macroscopic and the excitations behave as a gas of weakly interacting phonons~\cite{greywall1981thermal}.  
 
At 10-20 mK temperatures where the next generation of $^4$He Josephson junctions using arrays of $\sim$ 1 nm apertures is expected to operate, we can then identify three sources of dissipation that limit the overall $Q$-factor of the Helmholtz resonance, $Q_H$: i) three-phonon losses in the sensing loop~\cite{abraham1969sound}, ii) thermo-viscous losses produced by the junction~\cite{backhaus1997thermoviscoustheory}, and iii) mechanical dissipation of the diaphragm. Since the mass currents at the Helmholtz frequency have wavelengths comparable to the sensing loop dimensions, we treat the loop as an acoustic transmission line with a finite attenuation per unit length, similar to an electrical transmission line~\cite{pozar2021microwave}. 
Assuming that sensing loop losses can be completely described by three-phonon processes, as one of us has found with kilohertz acoustic resonator experiments~\cite{de2017ultra}, 
we find that a sensing loop with length $l$ and cross-sectional area $a_l$ can be described by a lumped circuit model with inductance $L_l = \frac{l}{\rho a_l}$ and resistance $R_l \propto T^4$ (Appendix \ref{app:B}).

Transport through the junction is pure, zero-entropy condensate flow, which changes the local entropy density and results in heat flow~\cite{backhaus1997thermoviscoustheory}.  This produces a temperature difference, $\Delta T$, due to the finite thermal conductivity and results in a chemical potential across the junction, $\Delta\mu/\rho = s \Delta T/\rho$, where $s$ is the entropy per volume in the 
superfluid. This chemical potential is in-phase with the current and acts to oppose the flow, realizing a resistive element. Assuming that thermalization occurs via transport of thermal (short-wavelength) phonons through the helium sensing loop, which scatter diffusely off the tube walls~\cite{greywall1981thermal}, we find $R_J \propto T^4$ (Appendix \ref{app:B}).

Losses due to mechanical friction in the motion of the diaphragm will also produce thermal noise.  These losses can be estimated from the microphonic motion of the diaphragm measured without helium in the circuit, yielding a temperature-independent effective resistance $R_d$. Taken together, these three sources of loss result in an overall $Q$-factor for the Helmholtz oscillator given by
\begin{equation}
    Q_H = \frac{\omega_H (L_J \parallel L_l)}{R_d + Re[Z_J \parallel Z_l]},
\end{equation}
where $\omega_H$ is the Helmholtz resonant frequency, $\parallel$ denotes the parallel combination of circuit elements and $Z (\omega, T) = i \omega L + R(T)$ are the complex impedances of the junction and sensing loop, respectively.
The thermal noise spectral density [rad/s/$\sqrt{\mathrm{Hz}}$]  is then
\begin{equation} \label{eqn:sensitivity}
    \sqrt{S_\Omega(\omega)} = \frac{1}{A}\sqrt{\frac{k_B T}{ Q_H}\frac{L_{J}(0)}{\omega_{oo}} }\cdot\frac{\epsilon{(\phi_0,\beta)}}{\phi_A}, 
\end{equation}
where $\omega_{oo}=(L_J(0) C_d)^{-1/2}$, $\epsilon{(\phi_0,\beta)}=\beta(\cos{\phi_0}+1/\beta)^{5/4}/\sin{\phi_0}$, $\beta = L_l/L_J(0)$, and $\phi_A$ is the amplitude of driven phase oscillations across the junction~\cite{avenel2004superfluid,chui2005frequency} (see also Appendix \ref{app:B}). The rotational sensitivity achievable in a measurement time $T_{meas}$ is then $\Delta \Omega = \sqrt{S_\Omega(\omega))}/\sqrt{T_{meas}}$.

Using the relation between phase and proper time, $\Delta \phi = mc^2 \delta \tau/\hbar$, and the relation between phase and angular velocity, $\Delta \phi = -\frac{2m}{\hbar} \vec{\Omega}_{tot} \cdot \vec{A}$, we can also find the spectral noise density on proper time in this setting,
\begin{equation}
    \sqrt{S_{\delta \tau}(\omega)} = \frac{2A}{c^2} \sqrt{S_\Omega(\omega)}
\end{equation}

\begin{figure} 
    \centering
    \includegraphics[width=0.5\textwidth]
    {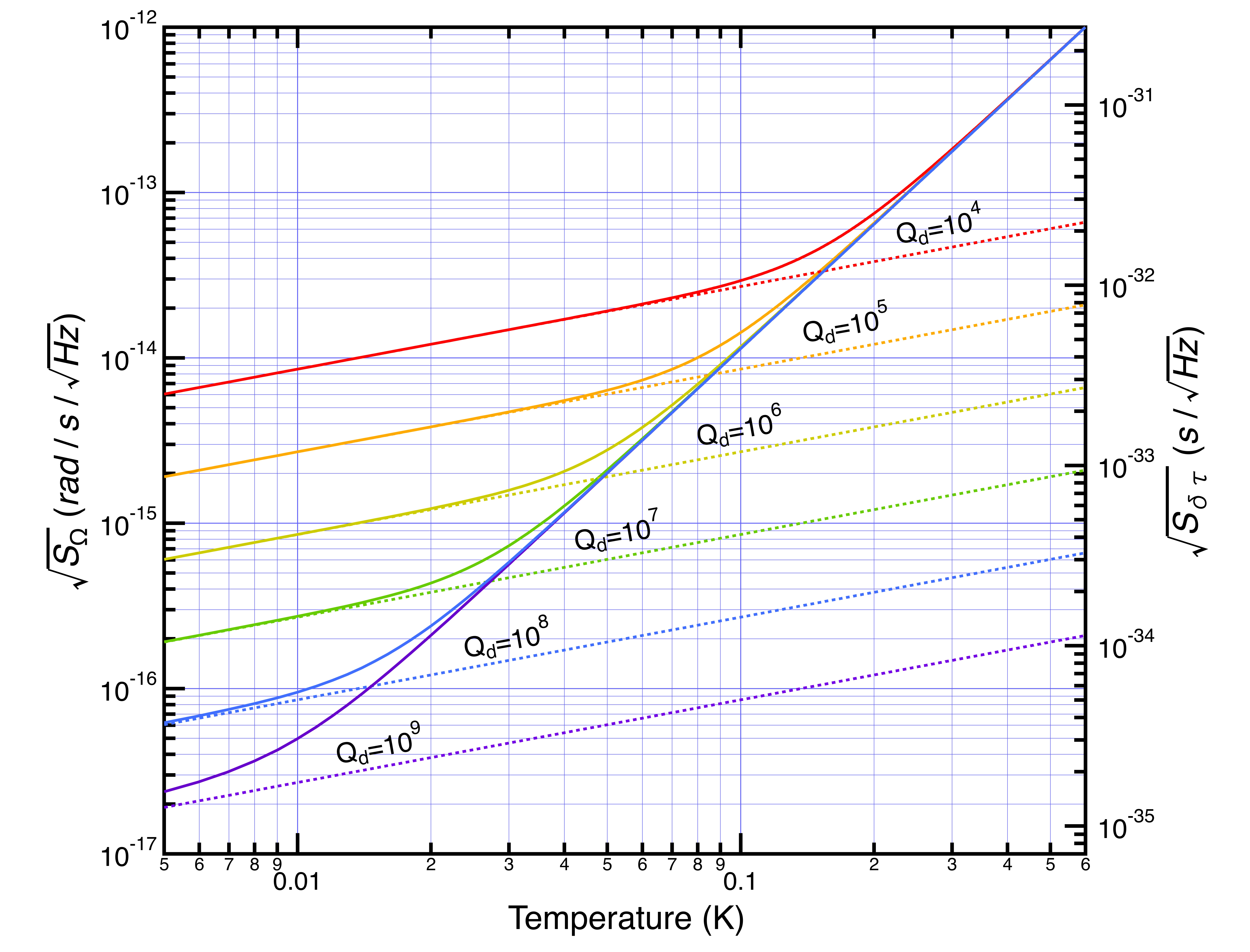}
    \caption{
    Noise spectral density on rotation rate (left vertical axis) and on proper time delay (right vertical axis) vs temperature for a single junction $^4$He SQUID, with diaphragm quality factors $Q_d=10^4-10^9$. The solid lines show the spectral density $\sqrt{S_\Omega}$ using dissipation from $R_l$, $R_J$, and $R_d$, while the dashed lines show the lower limit of   spectral density achieved from $R_d$ alone.
    The lowest noise spectral density measured previously with superfluid helium is $1\times 10^{-7}~\mathrm{rad/s}/\sqrt{\mathrm{Hz}}$ (for a $^3$He gyrometer~\cite{avenel2004superfluid}), and the best achieved with a ring laser is $2\times 10^{-12}~\mathrm{rad/s}/\sqrt{\mathrm{Hz}}$~\cite{di2024noise}.}
    \label{fig:sensitivity}
\end{figure}
Fig.~\ref{fig:sensitivity} shows the estimated noise spectral density on rotation measurements as a function of temperature, for a superfluid $^4$He gyrometer that is limited by the above three noise mechanisms, for several different values of the diaphragm quality factor $Q_d$. It 
is evident that the noise due to three-phonon and thermo-viscous effects drops rapidly with decreasing temperature, 
indicating that the low-temperature sensitivity 
will be ultimately limited by the mechanical dissipation of the diaphragm, $Q_d$.  Helmholtz resonators with quality factors of $Q_H = 70,000$ at 200 mK have been reported in previous experiments~\cite{backhaus1997thermoviscousexp}.  For the  estimates presented here, we assumed a sensing loop of radius 10 cm, formed by a single turn of 2 cm diameter copper tubing, with a diaphragm having a diameter of 16 mm, spring constant $10^4$ N/m, and a 3 kHz microphonic resonance.  The ratio $L_J/L_l=0.8$ is achieved with a junction critical mass current of $I_c=9.2\times 10^{-10}$ kg/s, 
compatible with critical mass currents measured in larger pores~\cite{hoskinson2006superfluid}. The Helmholtz resonant frequency is $\omega_H/2\pi = 101$ Hz. 

The position resolution required to observe thermal motion of the superfluid circuit with $Q_d=10^4$ at 10 mK is 110 fm/$\sqrt{\mathrm{Hz}}$, which can be achieved using standard membrane motion detection schemes~\cite{sato2009dc}.

Assuming a diaphragm quality factor $Q_d = 10^5$, and a fluid temperature of 10 mK, we estimate a thermal noise spectral density of $\sqrt{S_\Omega(\omega)} = 2.7\times 10^{-15}$ rad/s/$\sqrt{\mathrm{Hz}}$. This should be compared to the magnitude of the frame-dragging precession rate; choosing angles $\theta=90^\circ$ and $\chi=52.1^\circ$ (colatitude in Berkeley), the largest frame-dragging signal is obtained when $\psi=\theta-\chi$, giving $|\vec{\Omega}_{FD} \cdot \hat{n}|=2.9\times10^{-14}$ rad/s, where $\hat{n}$ is a unit vector normal to the loop surface. The required time for a frame-dragging measurement at the $0.2\%$  level is then $\sim35$ minutes. This is a significant improvement on the thermal noise spectral density predicted by~\cite{chui2005frequency} for $^3$He at $T=0.5 T_c$, which was $2\times10^{-13}$ rad/s/$\sqrt{\mathrm{Hz}}$. At that level of thermal noise, a $0.2\%$ measurement of frame-dragging would require 120 days of measurement.

Even better sensitivity could be achieved by using higher $Q$ materials for the diaphragm. Mechanical quality factors at these frequencies and temperatures have been demonstrated to exceed $10^9$ for crystals of Si, sapphire, and quartz~\cite{braginskiui1985systems,goryachev2012extremely,duffy1992acoustic}. Assuming a diaphragm $Q$ factor $Q_d=10^{9}$, leads to a predicted thermal noise floor at $10$ mK of $5 \times 10^{-17}$ rad/s/$\sqrt{\mathrm{Hz}}$. This would reduce the time for a frame-dragging measurement at the $0.2\%$ level to 0.7 s, which corresponds to rotational sensitivity of one revolution in $4\times 10^9$ yrs. This remarkable sensitivity also corresponds to a very small Josephson phase and an implied proper time difference of order 10$^{-35}$ s. between counter-propagating trajectories around the superfluid loop.

\emph{Discussion ---} Frame-dragging measurements to date relied on the accumulation of the effect over many orbital periods. Our proposal provides an Earth-based measurement, over seconds. The extremely low thermal noise enables frame-dragging measurements to the 0.2$\%$ level or better, providing strong motivation to pursue this approach. Moreover, any systematic errors in a superfluid gyrometer would likely be independent of those in previous, space-based measurements. 

The extreme sensitivity estimated here for $^4$He gyrometers at millikelvin temperatures demonstrates a key advantage over $^3$He devices, where the observed $Q_H$ has not exceeded 200, the limiting dissipation mechanisms are not yet clear, and the temperatures required are much lower~\cite{hook1995diaphragm,avenel2004superfluid}. This motivates the development of custom-designed nanomaterials, such as nanoporous molecular membranes~\cite{kissel2014nanoporous}, for use as $^4$He junctions at millikelvin temperatures. 
As is clear from Fig.~\ref{fig:sensitivity}, the thermal noise in millikelvin $^4$He devices will ultimately be limited by mechanical design and the ability to control mechanical and acoustic losses. The extraordinarily low level of noise shown in Fig. \ref{fig:sensitivity} will only be achieved if all other sources of random phases have been eliminated, 
including possible motion of trapped quantized vortices and effects of external vibrations. However, past superfluid gyrometer experiments have not shown evidence for noise deriving from trapped vorticity~\cite{schwab1997detection,avenel2004superfluid},
while the effects of external vibrations are expected to be mitigated by the vibrational isolation techniques developed for LIGO~\cite{matichard2015advanced} and RLGs~\cite{schreiber2023variations}. 

Let us briefly mention some possible advantages of superfluid devices for applications in geodesy and frame-dragging measurements, as compared to RLGs. The most sensitive RLG, GINGERINO, has demonstrated a noise level of $2\times 10^{-12}$ rad/s/$\sqrt{\mathrm{Hz}}$ and has sufficient stability to reach a sensitivity of $\Delta \Omega = 2\times 10^{-15}$ rad/s in $2\times 10^5$s of integration time~\cite{di2024noise}, which should be low enough to allow measurements of frame-dragging at the 10\% level. The GINGER collaboration aims to build three such RLGs with different orientations, so as to measure all components of the total rotation vector~\cite{altucci2023status}. By comparison, in this work we have estimated that for superfluid $^4$He gyrometers with nanoscale apertures a thermal noise floor on the order of $10^{-15}-10^{-17}$ rad/s/$\sqrt{\mathrm{Hz}}$ 
can be achieved. In $2\times 10^5$ s of measurement, this would give rotational sensitivity 
of $\Delta \Omega \sim 10^{-17}-10^{-19}$ rad/s.
We also note that while precision RLG experiments such as GINGER typically require an enclosed sensing area greater than 10 m$^2$, our proposed superfluid gyrometer has an area of just $3\times 10^{-2}~\mathrm{m^2}$. In addition,
the ability to reorient a helium gyrometer~\cite{simmonds2001quantum} removes the need for three independent gyrometers. 

We have calculated that, strikingly, the proper time difference probed by our superfluid gyrometer is at the order of $10^{-35}$ s. This is comparable to that required in proposals to detect the non-classicality of the gravitational field via entanglement through gravity~\cite{rovelli2018possibility}. It is interesting to investigate the potential of similar helium interferometry techniques for applications in table-top quantum gravity tests~\cite{lantano2024low,christodoulou2019possibility,christodoulou2022experiment}.

In summary, the development of millikelvin $^4$He Josephson junctions provides the opportunity for an Earth-based frame-dragging measurement in a well-controlled environment, with a clear potential to significantly improve the current precision to which this effect has been measured. The practical advantage of using such quantum devices as compact, orientable gyrometers, makes them an appealing complementary tool to other leading gyroscopic technologies for measurement of minute relativistic gravitational effects. Finally, we note that such a measurement would demonstrate a truly general-relativistic effect in a macroscopic quantum coherent system.

\begin{acknowledgments}
It is our pleasure to acknowledge useful conversations with Richard Packard, Rafaelle Silvestri, Markus Aspelmeyer, Holger Mueller, Kranthi Mandadapu, Francis Headley, Yixin Xiao and William Loinaz.
KIE and KBW acknowledge support from the NSF QLCI program through grant number OMA-2016245, and thank the Institute for Quantum Optics and Quantum Information (IQOQI) in Vienna for hosting them in spring 2023 where this work was started.
KCS acknowledges support from the NSF QuSeC-TAQS program through grant number 2326801 and from the NSF DMR program through grant number 2103425. MC acknowledges the support of the ID\# 62312 grant from the John Templeton Foundation (The Quantum Information Structure of Spacetime Project, QISS).
\end{acknowledgments}

\bibliography{bibliography}

\begin{thebibliography}{62}
\bibitem [{not()}]{noteSagnac}%
  \BibitemOpen
  \href@noop {} {}\bibinfo {note} {The Sagnac effect would also appear as a contribution to the off-diagonal metric elements $g_{0i}$ in a frame that is comoving and corotating with the gyrometer}\BibitemShut {NoStop}%
\end{thebibliography}
\bibliographystyle{apsrev4-2.bst}

\appendix

\section{Lense-Thirring metric in comoving frame and superfluid velocity} 
\label{app:A}
The metric given in Ref.~\cite{ashby1990geodetic} in the comoving frame can be written as
\begin{equation}
    \label{metric_2}
    \begin{aligned}
        &g_{00} = -1+\frac{2 \vec{R} \cdot \vec{a}}{c^2} \, , \qquad g_{ij} = \delta_{ij} \, , \\ 
        &\vec{g} = -\frac{1}{c}(\vec{\Omega}_{FD} + \vec{\Omega}_{G}+\vec{\Omega}_{T}) \times \vec{R} + K \vec{R} \, ,
    \end{aligned}
\end{equation}
where
\begin{equation} \label{eq:omega_1}
\begin{aligned}
    &\vec{\Omega}_{FD} =  \frac{(1+\gamma+\frac{1}{4}\alpha_1)}{c^2} (\nabla \times \vec{H}) \\ 
    &\vec{\Omega}_G = \frac{(2\gamma+1)}{2c^2} (\vec{v} \times \nabla U) \, , \\ 
    &\vec{\Omega}_T = \frac{1}{2c^2} (\vec{a} \times \vec{v}) \, ,  \qquad K = \frac{\gamma}{c^3} (\vec{v} \cdot \nabla U)
\end{aligned} 
\end{equation}
are the frame-dragging, geodetic and Thomas precession rates, expressed in terms of the two functions,
\begin{equation}
    U=U(r)=\frac{GM_\oplus}{r}
\end{equation}
and
\begin{equation}
    \vec{H}=\vec{H}(\vec{r})=\frac{GM_\oplus R_\oplus^2}{5r^3} (\vec{\Omega}_\oplus \times \vec{r}).
\end{equation}
The coordinates $\vec{r}$ are those of the parameterized post-Newtonian (PPN) frame, in which the solar system is approximately at rest and the PPN coordinates are nearly globally Lorentz~\cite{ashby1990geodetic,will2014confrontation,misner1973gravitation}. The velocity, $\vec{v}$, of the gyrometer is also referred to the PPN frame. All gradients in \eqref{eq:omega_1} are taken with respect to the PPN coordinates and evaluated at the origin of the comoving frame~\cite{ashby1990geodetic}. 

In addition to the PPN parameter $\gamma$ included in Ref.~\cite{ashby1990geodetic}, we have additionally included $\alpha_1$, for consistency with the PPN literature~\cite{will2014confrontation}, although we have not included preferred frame terms, which are sometimes included~\cite{will2014confrontation,bosi2011measuring}.

The precession rates can be written in a more illuminating form by evaluating the gradients and curls in \eqref{eq:omega_1}, 
\begin{equation} \label{omega}
\begin{aligned}
    &\vec{\Omega}_{FD} =  \frac{(1+\gamma+\frac{1}{4}\alpha_1)GM_\oplus R_\oplus^2}{5 c^2 r^3} \left(\frac{3(\vec{\Omega}_\oplus \cdot \vec{r}) \vec{r}}{r^2} - \vec{\Omega}_\oplus \right) \, , \\
    &\vec{\Omega}_G = \frac{(2\gamma + 1)}{2} \frac{GM_\oplus}{c^2 r^3} \left( \vec{r} \times \vec{v} \right) \, , \qquad \vec{\Omega}_T = \frac{1}{2c^2} (\vec{a} \times \vec{v}) \, .
\end{aligned} 
\end{equation}

Recall that $\vec{a}$ is the acceleration of the gyrometer away from geodesic motion, so for a gyrometer on Earth it is the sum of the accelerations due to the normal force counterbalancing gravity and due to the Earth's rotation~\cite{schiff1960possible},
\begin{equation}
    \vec{a} = \frac{GM_\oplus}{r^3} \vec{r} + \frac{d\vec{v}}{dt} \, .
\end{equation}
For a gyrometer in an Earth-bound laboratory we have $\vec{v}=\vec{\Omega}_\oplus \times \vec{R}_\oplus$ and $\frac{d\vec{v}}{dt}=\vec{\Omega}_\oplus \times \vec{v}$. The form of the Thomas phase given in \eqref{eq:phases} follows directly from these observations.

Using the metric $g_{\mu \nu}$ in the comoving frame in \eqref{metric}, we can expand the point particle Lagrangian, $L = -m (-g_{\mu\nu} \dot{x}^\mu \dot{x}^\nu )^{1/2}$, in the weak-field limit, as
\begin{equation}
    L \approx -m - \frac{m}{2}\left( -h_{00} - 2g_{0i} \dot{x}^i - \dot{x}^i)^2  \right) \, .
\end{equation}

Then the canonical momentum conjugate to $x^i$ is $p_i = \frac{\partial L}{\partial \dot{x}^i} = m g_{0i} + m \dot{x}_i$. Solving for $\dot{x}_i$ in terms of $p_i$ and substituting back into the Lagrangian, we find the Hamiltonian $H = p_i \dot{x}^i - L$ to be given by
\begin{equation} \label{hamiltonian}
    H = \frac{1}{2m}\left(\vec{p} - mc\vec{g} \right)^2 - \frac{1}{2} mch_{00} \,.
\end{equation}
The fact that we can write the Hamiltonian in this simple form is a special property of the comoving reference frame that we have chosen to work in; in particular, it is due to the fact that $g_{ij}=\delta_{ij}$ in this frame. 

Notice that this Hamiltonian has the same form as the Hamiltonian for a charge $q$ moving in in an electromagnetic field with scalar potential $\phi$ and vector potential $\vec{A}$, if we identify $q \leftrightarrow m$, $\vec{A} \leftrightarrow c \vec{g}$ and $\phi \leftrightarrow \frac{1}{2} c h_{00}$. Therefore, we can follow the logic used in Ref.~\cite{Feynman:1494701} for superconductors and apply it to our Hamiltonian to find the superfluid current
\begin{equation}
    \vec{J} = \frac{1}{2m} \left(\psi^* \left[\frac{\hbar}{i} \nabla - mc\vec{g} \right] \psi + \psi \left[\frac{\hbar}{i}\nabla - mc\vec{g}\right]^* \psi^* \right) \, .
\end{equation}
 Substituting in the superfluid wavefunction, $\psi=\sqrt{\rho} e^{i\phi}$, and identifying $\vec{J}=\vec{v}_s \rho$ gives the superfluid velocity $\vec{v}_s = \frac{\hbar}{m} \nabla \phi - c \vec{g}$.
\label{sec:appendixA}

\twocolumngrid

\section{Details on noise estimation}
\label{app:B}
To estimate the rotation sensitivity of a single junction $^4$He gyrometer limited by thermal fluctuations, we follow the analysis in Ref.~\cite{avenel2004superfluid} and \cite{chui2005frequency}, with circuit variables of $I$ with units of kg/s, and potential given by $\Delta \mu / m_4 = \Delta P/\rho-s\Delta T/\rho$ with units of J/kg, where $\Delta P$ and $\Delta T$, and $s$ are the pressure and temperature differences and the entropy per volume. 

The Josephson effects produces non-linear dynamics.  To avoid the sensitivity limitations which would be required to stay within the linear regime for a continuous drive as in Ref.~\cite{avenel2004superfluid}, we consider the procedure described in Ref.~\cite{chui2005frequency} where the system evolves freely after an impulsive drive, and then period is measured.

Since the Josephson inductance of the junction is phase-dependent through $L_J(\phi_J)=\kappa_4  /(2\pi I_c \cos{\phi_J})$, the period of oscillation of the circuit shown in Fig.~\ref{fig:gyroscope} is also perturbed by changes in $\phi_J$: $\omega_H(\phi_J) = 1/\sqrt{(L_l \parallel L_J(\phi_J)) C_d}$. To detect changes in the quantum phase due to rotation, the resonator diaphragm is given an impulse, producing an oscillating mass current and resulting quantum phase oscillation across the junction with amplitude $\phi_A$ about the static phase of $\phi_0$: $\phi_J(t)=\phi_0+\phi_A(t)$.   Small, quasi-static changes in the quantum phase due to rotation will perturb $\phi_0$, and result in a change in the period of oscillation. 

Noise enters the circuit through dissipation, where the power spectral density of fluctuations in the chemical potential is given by: $S_{\Delta P/\rho} = 4 k_B T R$, where $R = (\Delta P/\rho) / I$  is any hydrodynamic resistance in the circuit which determines the quality factor of the resonance, $Q_H$.  With these assumptions, we obtain the result in~\eqref{eqn:sensitivity} by adapting the analysis of Ref.~\cite{chui2005frequency} to the case of pure superfluid $^4$He, applicable for $^4$He below $\sim$~0.6 K~\cite{guenault2019probing} (compare to Eqn.~11 in~\cite{chui2005frequency}, noting that, in the pure superfluid limit, the frequency $f_{oo}$ in ~\cite{chui2005frequency} is related to $\omega_{oo}$ of \eqref{eqn:sensitivity}  by $\omega_{oo}=2\pi f_{oo}$). It is apparent that the gyrometer noise is minimized by reducing the temperature and increasing the quality factor.  Cooling $^4$He to 10 mK is possible with a standard dilution refrigerator.

We find three sources of dissipation in the circuit that will limit $Q_H$: losses in the sensing loop $R_l$, in the junction $R_J$, and in the diaphragm $R_d$. A full analysis of each loss mechanism will be published elsewhere; here we summarize the key assumptions and results.

The hydrodynamic resistance in the sensing loop can be understood by treating the loop as an acoustic transmission line, similar to the analysis of electrical lines~\cite{pozar2021microwave}.  In the limit where the $\omega_H$ is below the first acoustic resonance of the line, the effective resistance is given by $R_l = \mathcal{Z}_0 \alpha l$, where $\mathcal{Z}_0 = (\rho^3 a_l^2 \beta)^{-1/2}$ is the characteristic acoustic impedance of the sensing loop, $\beta$ is the compressibility of superfluid helium,
$a_l$ is the cross-sectional area of the line and $\alpha$ is the attenuation per unit length, which can arise from intrinsic three-phonon loss in the helium, boundary losses into the containing structure, or the viscosity of dilute helium-3 impurities.  Experiments with audio frequency superfluid acoustic resonators~\cite{de2017ultra} show that the three-phonon losses dominate after the losses into the boundary are minimized.  At 100 Hz, losses due to helium-3 impurities are expected to be minimal in the temperature range 10 mK-600 mK.  Assuming that $\alpha$ is entirely due to three-phonon losses, we find:
\begin{equation} \label{hydro}
    R_l = \sqrt{\frac{1}{\rho^3 a_l^2 \beta}} \frac{\pi^3}{60} \frac{\left(G+1\right)^{2} }{\rho \hbar^3 c_{4}^{6}} \left(k_{B}T\right)^{4} l \omega,
\end{equation}
with $G=2.84$ is the Gr\"uneisen parameter, $c_4$ the speed of sound in superfluid $^4$He and $\omega$ the frequency of the acoustic mode~\cite{abraham1969sound}.

$R_J$ is due to thermo-viscosity~\cite{backhaus1997thermoviscoustheory,backhaus1997thermoviscousexp}: mass currents through the Josephson junction are zero-entropy superflow which change the density of thermal excitations around the junction.  This leads to heat flow, resulting in temperature differences and a chemical potential across the junction opposing the flow. In the temperature range considered here, heat moves primarily through phonons and estimates suggest that the  transport is through the fluid in the sensing loop, rather than through the containing structure, due to Kapitza boundary resistance~\cite{pobell2007matter}.  Due to the short wavelength of the thermal phonons (400 nm--10 nm in this temperature range), the mean free path of the phonons is typically found to be limited by the diameter of the sensing loop tubing~\cite{greywall1981thermal}. This yields
\begin{equation}
    R_{J} = \sqrt{\frac{\pi}{a_l^3}} \frac{ l~s~T}{2\rho^2 c_4} .
\end{equation}

Lastly, we consider the dissipation due to the mechanical losses in the motion of the diaphragm.  Treating the diaphragm as a damped simple harmonic oscillator with natural frequency $\omega_d$ and quality factor $Q_d$ (measured with an empty cell), we find
\begin{equation}
    R_d = \frac{1}{C_d \omega_d Q_d},
\end{equation}
 where $C_d=(a_d\rho)^2/k_d$ is the hydrodynamic capacitance of the diaphragm, 
 given in terms of the diaphragm area $a_d$ and spring constant $k_d$.

For the plot in Fig.~\ref{fig:sensitivity}, we assume $\beta = L_l/L_J(0)=0.8$, $\phi_0=2.3$ and $\phi_A=0.2$ to assure stability and maximize the sensitivity to small shifts in $\phi_0$ due to rotation~\cite{chui2005frequency}; $a_d=2\cdot 10^{-4}~\mathrm{m}^2$,  $k_d=10^4~\mathrm{N/m}^2$, $\omega_d/(2\pi) = 3000~\mathrm{Hz}$; sensing loop tubing area and loop area: $a_l=3\cdot 10^{-4}~\mathrm{m^2}$, $A=3\cdot 10^{-2}~\mathrm{m^2}$.  The junction critical current assumed is $I_c=9.2\cdot 10^{-10}$ kg/s, which corresponds to a superfluid velocity of $2.0$ m/s over a 2 $\mu$m diameter junction, and is comparable to critical velocities observed in micron-scale apertures~\cite{davis1992evidence}. The actual critical current density has yet to be determined for nanoporous material such as poly-fantrip~\cite{kissel2014nanoporous} and will depend on the degree of suppression of the superfluid density in the pores, and the total effective area of the pores in the membrane.

Fig.~\ref{fig:gyroscope}b shows that the diaphragm is in series with the rest of the circuit, which consists of a Josephson junction and sensing loop in parallel with each other.

Assuming that the dissipative impedance in both junction and loop is small compared to the impedance of the reactive elements, i.e., $R_J, R_l << \omega L_J, \omega L_l$, the full circuit will then behave as an RLC series circuit with inductance $L_J(\phi_0) \parallel L_l = (1/L_J(\phi_0) + 1/L_l)^{-1}$, resistance $R_{tot} = R_d + Re[Z_J \parallel Z_l]$ and capacitance $C_d$. 

\end{document}